\documentclass[letterpaper, 10 pt, conference,table]{ieeeconf}  

\IEEEoverridecommandlockouts                              

\usepackage{cite}
\usepackage{comment}
\usepackage{todonotes}
\usepackage{amsmath,amssymb,amsfonts}
\usepackage{algorithm}
\usepackage[noend]{algpseudocode}
\usepackage{booktabs}
\usepackage{adjustbox}
\usepackage{graphicx}
\usepackage{textcomp}
\usepackage{xcolor}
\usepackage{paralist}
\usepackage{booktabs, makecell, tabularx}
\usepackage{caption}
\usepackage{changepage}
\usepackage{multirow}
\usepackage{microtype}
\usepackage{siunitx}

\title{\LARGE \bf
ErgoTac-Belt: Anticipatory Vibrotactile Feedback to Lead \\Centre of Pressure during Walking}

\author{Marta Lorenzini$^{1}$, Juan M. Gandarias$^{1}$, Luca Fortini$^{1}$, Wansoo Kim$^{2}$ and Arash Ajoudani$^{1}$
\thanks{*This work was supported in part by the ERC-StG Ergo-Lean (Grant Agreement No.850932), in part by the European Union’s Horizon 2020 research and innovation programme under Grant Agreement No. 871237 (SOPHIA).}
\thanks{$^{1}$The authors are with the Human-Robot Interfaces and Physical Interaction (HRI$^{2}$) Lab, Istituto Italiano di Tecnologia, Genoa, Italy.
        ({\tt\small marta.lorenzini@iit.it})}%
\thanks{$^{2}$W. Kim is with Robotics Department,  Hanyang University ERICA, Republic of Korea}%
}

\begin{document}

\maketitle
\thispagestyle{empty}
\pagestyle{empty}

\begin{abstract}
Balance and gait disorders are the second leading cause of falls, which, along with consequent injuries, are reported as major public health problems all over the world. For patients who do not require mechanical support, vibrotactile feedback interfaces have proven to be a successful approach in restoring balance. Most of the existing strategies assess trunk or head tilt and velocity or plantar forces, and are limited to the analysis of stance. On the other hand, central to balance control is the need to maintain the body's centre of pressure (CoP) within feasible limits of the support polygon (SP), as in standing, or on track to a new SP, as in walking. Hence, this paper proposes an exploratory study to investigate whether vibrotactile feedback can be employed to lead human CoP during walking. The ErgoTac-Belt vibrotactile device is introduced to instruct the users about the direction to take, both in the antero-posterior and medio-lateral axes. An anticipatory strategy is adopted here, to give the users enough time to react to the stimuli. Experiments on ten healthy subjects demonstrated the promising capability of the proposed device to guide the users' CoP along a predefined reference path, with similar performance as the one achieved with visual feedback. Future developments will investigate our strategy and device in guiding the CoP of elderly or individuals with vestibular impairments, who may not be aware of or, able to figure out, a safe and ergonomic CoP path.
\end{abstract}

\section{Introduction}
\label{sec:introduction}

Postural control is a complex motor skill that requires the interaction of the muscular, nervous, and sensory systems. Most notably, the human central nervous system must process and integrate concurrent feedback signals from the visual, vestibular, and somatosensory channels~\cite{horak1996postural}. If one or more of these channels are impaired or even absent, postural control is compromised, and the risk of falling perilously increases~\cite{melzer2004postural}. Indeed, balance and gait disorders are the second leading cause of falls, just coming after accidents~\cite{rubenstein2002epidemiology}. 
Falls and consequent injuries, in turn, have been and are reported as a major public health problem all over the world~\cite{tinetti2003preventing}. 
Therefore, a lot of effort has been put recently in the development of devices and strategies to improve postural control and therefore to avoid falls \cite{van2019robotic,di2015fall,ruiz2021improving}.

For patients who do not require mechanical support, feedback interfaces have proven to be a promising approach in fall prevention~\cite{sienko2017role}. Visual, auditory, electrotactile, and vibrotactile feedback interfaces can provide additional sensory information and can be used to improve postural stability, decreasing the postural sway and even the risk of falling~\cite{wall2001balance}.  
In particular, vibrotactile devices seem to show the most promise, since they are wearable, unobtrusive, and do not interfere with the human sensory systems (e.g., visual and auditory)\cite{ma2016balance}.
Plenty of literature exists on vibrotactile feedback strategies to enhance postural stability and various possibilities have been explored. Above all, several strategies were proposed to discriminate between instability and stability, and thus trigger the vibrotactile stimuli.
In most of the studies, the authors assessed the trunk tilt or velocity~\cite{tannert2021immediate,ballardini2020vibrotactile,xu2017configurable,nanhoe2012effects,sienko2012biofeedback,haggerty2012effects,alahakone2010real,kentala2003reduction,verhoeff2009effects} or head tilt~\cite{janssen2010salient,wall2001balance} collected through IMUs or simple accelerometers. 
Alternatively, some researchers presented methods based on plantar forces~\cite{ma2015vibrotactile,ma2017wearable,afzal2015portable,crea2014providing}, which require the subjects to wear pressure-sensitive insoles or force-sensitive resistors (FSRs) embedded in the shoes. 

\begin{figure}
    \centering
    \includegraphics[width=0.85\columnwidth]{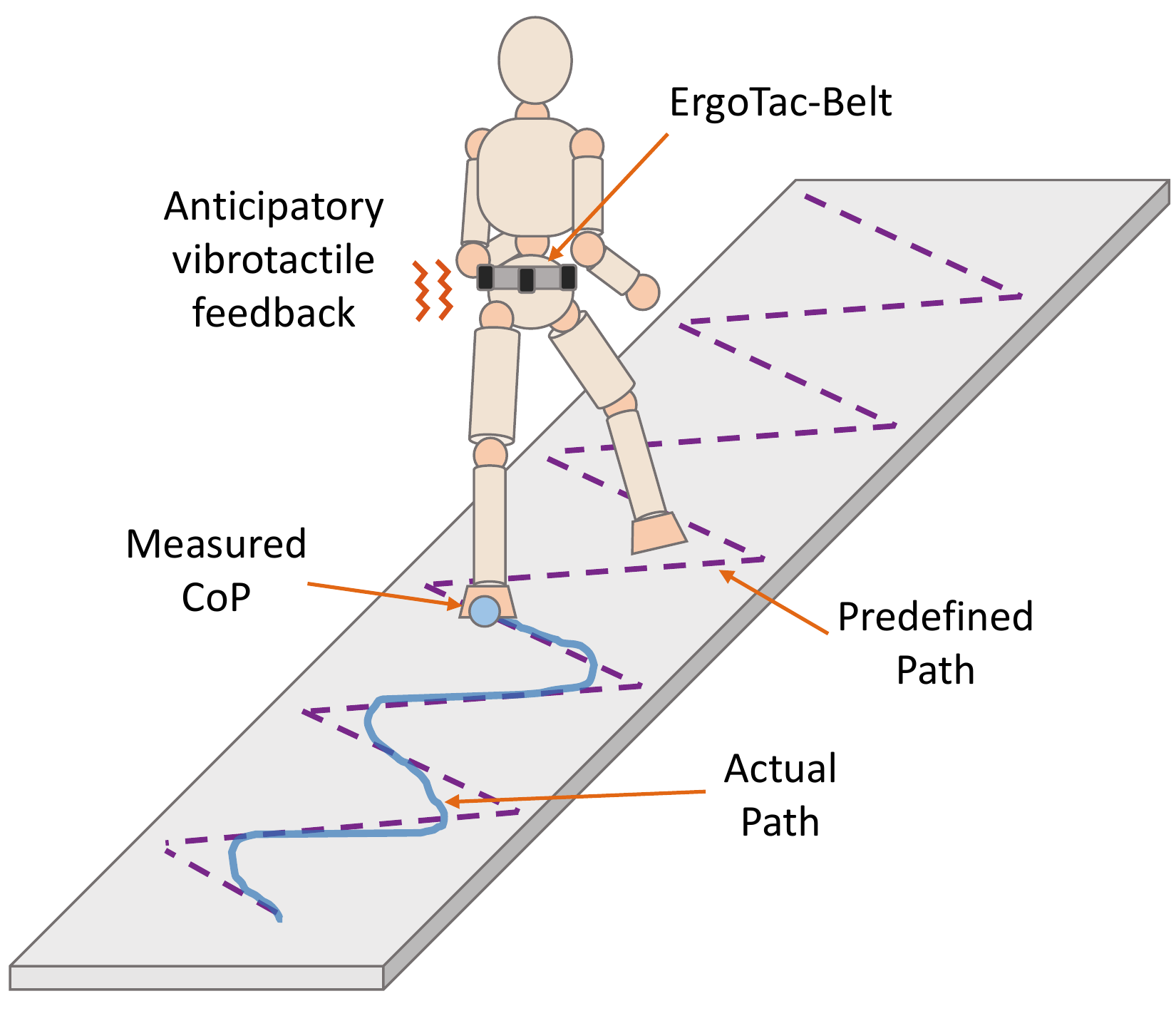}
    \caption{An illustration of a human being driven by the vibrotactile device ErgoTac-Belt. The device provides anticipatory vibrotactile feedback from the displacement of the measured Centre of Pressure (CoP) wrt a predefined CoP path. }
    \label{fig:digest}
    \vspace{-5 mm}
\end{figure}

On the other hand, central to balance control is the need to maintain the body’s centre of
mass (CoM) within manageable limits of the support polygon (SP) as in standing, or on track to a new SP, as in walking or running~\cite{winter1995abc}. Hence, the human whole-body centre of pressure (CoP), which is strictly related to the CoM in quasi-static conditions, might be the most suitable parameter for characterising balancing ability. 
Indeed, several studies considered the CoM/CoP as a measure of general postural stability to validate the proposed strategies~\cite{tannert2021immediate,ma2015vibrotactile,ma2017wearable,sienko2012biofeedback,lee2012directional}. But few of them directly employ CoM/CoP to recognise instability and thus activate the vibrotactile feedback~\cite{vuillerme2007controlling,fani2021multi}.
Furthermore, vibrotactile devices have been considered far and wide to improve postural stability during standing~\cite{wall2001balance,kentala2003reduction,vuillerme2007controlling,alahakone2010real,haggerty2012effects,sienko2012biofeedback,lee2012directional,ma2015vibrotactile,kinnaird2016effects,xu2017configurable,ma2017wearable,ballardini2020vibrotactile,tannert2021immediate,fani2021multi} while, to the best of our knowledge, not many authors examined the effect of vibrotactile stimuli on balancing while walking~\cite{verhoeff2009effects,nanhoe2012effects}. In these latter studies, an investigation was conducted on the effect of feedback on trunk sway in healthy young or older subjects during dual task~\cite{verhoeff2009effects} or in Parkinson's disease patients~\cite{nanhoe2012effects}.

In view of the above, this paper proposes an exploratory study to investigate whether vibrotactile feedback can be employed to lead human CoP during walking (see Fig. \ref{fig:digest}). Specifically, we implement an anticipatory strategy to provide the subjects with a vibrotactile input to follow a certain direction slightly in advance, to give them enough time to react to the stimulus. 
We hypothesise that if we can lead the human CoP to follow a predefined reference path through vibrotactile guidance, we may improve balancing even in more complex tasks than standing. Hence, we first consider here healthy subjects and we attempt to guide, during walking, their CoP along a reference path, which may differ from their usual gait, by using a vibrotactile feedback device. If a specific gait pattern can be imposed on healthy subjects through an anticipatory use of CoP-based vibrotactile feedback, a normal one may be restored in individuals with vestibular deficits or elderly with diminished balancing capacity. 

\section{Material and Methods}
\label{sec:methods}

This section first presents the vibrotactile feedback device employed in this paper. Next, we illustrate the anticipatory strategy developed to lead the human CoP path during walking via vibrotactile feedback.

\subsection{ErgoTac-Belt: a vibrotactile feedback belt}
\begin{figure}
    \centering
    \includegraphics[width = 0.95\columnwidth]{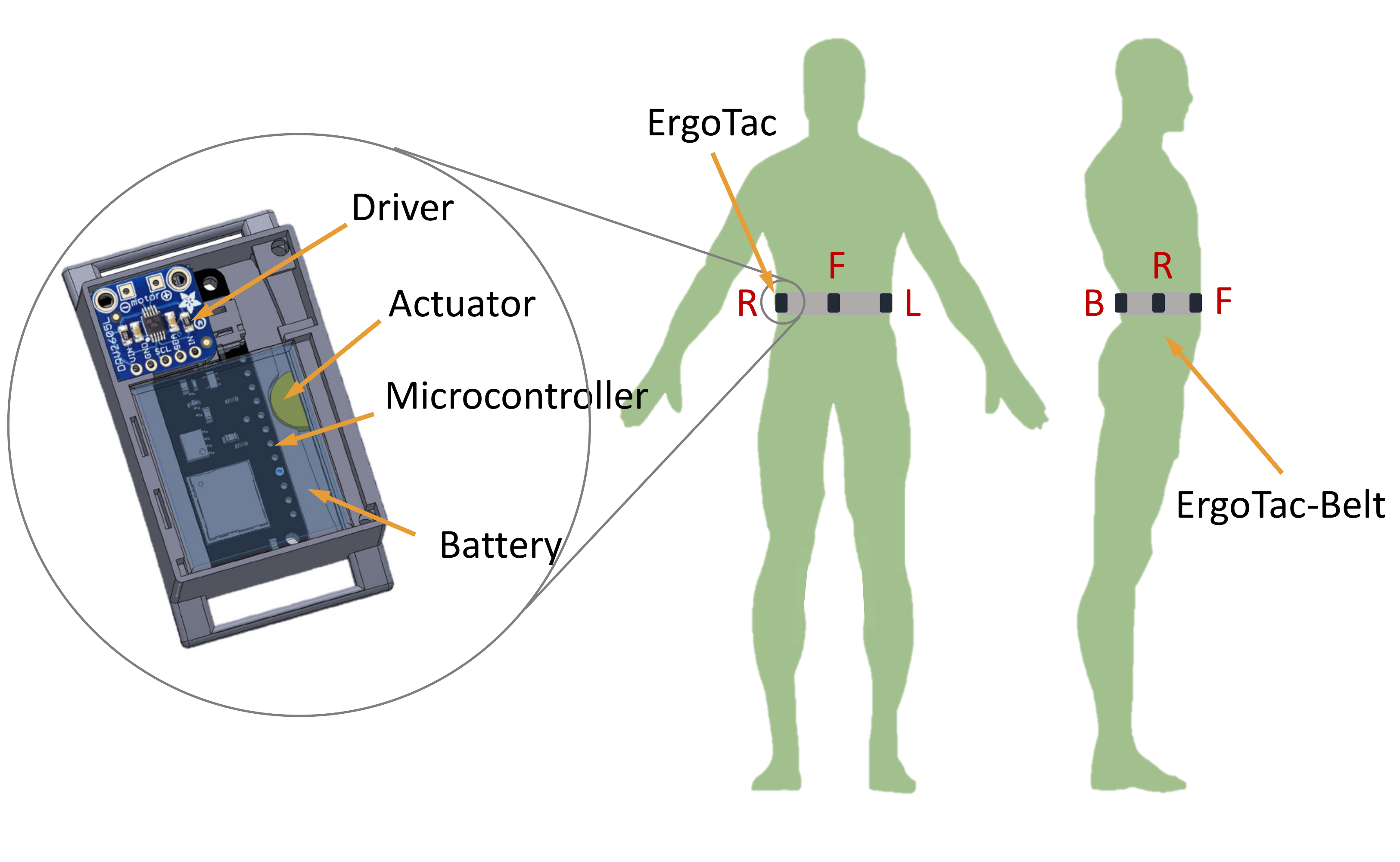}
    \caption{\small An illustration of ErgoTac-Belt, the vibrotactile device employed in this work. ErgoTac-Belt is composed of four ErgoTac devices located at the front (F), right (R), left (L), and back (B) of the user's torso, at L5 level.}
    \label{fig:ErgoTac-Belt}
    \vspace{-5 mm}
\end{figure}

When dealing with vibrotactile stimulation, some fundamental aspects must be considered:
\begin{itemize}
    \item \textit{Intensity of the vibrations.} Vibration intensity was observed to have a considerable effect on user experience. High intensities are easier to be detected~\cite{Shull2015}, but vibration-associated injuries may arise when increasing the intensity and duration of the vibration exposure~\cite{tan2019user}. Since we expect future users to wear the vibrotactile device for a prolonged time, low-intensity vibrations must be selected to maximise usability.
    \item \textit{Duration of the vibrations.} To achieve a clear perception, the duration of the vibrotactile stimulation should be over $100$ ms~\cite{filosa2018new,crea2017time}. However, a longer duration is needed when lower intensity is used~\cite{Li2015}. Hence, vibrations duration was selected to be to $400$ ms, as in~\cite{kim2021directional}, to compensate for the low intensity. 
    \item \textit{Position of vibrotactile units.} Several studies have been conducted to determine the best location for vibrotactile feedback units depending on the desired objective. For quiet and perturbed stance, torso-based vibrotactile feedback has been shown to improve postural performance in healthy young and older adults and individuals with balance impairments~\cite{ma2016balance}. Therefore, in this study, vibrotactile units are placed on the torso at the L5 vertebra level (i.e., the CoM area).
    \item \textit{Number of vibrotactile units.} A high number of vibration actuators may force the user to process a redundant set of signals, resulting in non-intuitive or complex patterns of somatosensory stimuli~\cite{brewster2004tactons}. As in~\cite{ballardini2020vibrotactile}, we then provide the subjects with a reduced number of vibrotactile units, i.e., two for the anteroposterior (AP) direction, placed front and back, and two for the mediolateral (ML) direction, placed left and right. 
    \item \textit{Instructional cue.} Different ways of encoding vibrotactile input can be instructed to the users. They may be asked to move in the opposite direction of the stimulus (repulsive cue) or the same direction (attractive cue). Researchers have extensively investigated both possibilities. As in our previous works~\cite{kim2018ergotac,kim2021directional}, we selected the repulsive strategy.  
\end{itemize}

According to the aforementioned considerations, a vibrotactile belt, namely ErgoTac-Belt is designed to provide vibrotactile feedback to the user. The device is composed of four vibrotactile units located at the front~(F), right~(R), left~(L), and back~(B) of the person, at the L5 level. This way, the ErgoTac-Belt can provide vibrotactile feedback to guide the person towards different locations on the plane based on the actual CoP and a predefined CoP path. An illustration of the device and how is worn by the user is depicted in Fig.~\ref{fig:ErgoTac-Belt}.
The vibrotactile units that constitute the ErgoTac-Belt are feedback devices called ErgoTac, which we presented in~\cite{kim2018ergotac} and further extended in~\cite{kim2021directional}. A detailed CAD render of ErgoTac along with its components is illustrated zoomed out in Fig.~\ref{fig:ErgoTac-Belt}. 
ErgoTac was designed as a wireless vibrotactile device to be worn at different human body segments. It was conceived to warn the user when exceeding different ergonomic indicators in order to adjust their posture into a more ergonomic one when developing physical tasks.
Due to its dimension ($68.1$ mm $\times$ $37.0$ mm $\times$ $17.3$ mm), weight ($28$ g) and communication protocol (multi-point connection via Bluetooth low energy at $2.4$ GHz), ErgoTac is ideal for applications in which wearability is required. 

\subsection{Feedback strategy to drive the centre of pressure}

This paper aims to exploit the ErgoTac-Belts vibrotactile feedback to induce the users to follow a predefined reference CoP path. Basically, at each instant, they are required to match their current CoP, which can be measured with an external sensor system (e.g., a force plate) with a reference CoP. The vibrotactile input from ErgoTac-Belt can provide the direction they have to take, both in the AP or ML axes, to accomplish this task. Specifically, we implement an anticipatory feedback algorithm to provide stimuli slightly in advance to give the users enough time to react and take the correct direction.
Algorithm~\ref{alg:feedback} presents the pseudocode of the feedback strategy adopted. 
Prior to the task, a threshold $th_{\text{CoP}}$ and an anticipatory time interval $t_{A}$ are selected. Then, the task starts, and a while loop is executed until the reference CoP path is terminated. At each step $t$ of the loop, the difference between the reference CoP $(\bar x, \bar y)$ and the measured CoP $(x,y)$, both in the AP and ML axes, is computed as
\begin{equation}
    \begin{cases}
        |{\delta x}_{\text{CoP}}|^{A} = |{\bar x(t + t_{A})}_{\text{CoP}}-{x(t)}_{\text{CoP}}| & \text{for AP} \\
        |{\delta y}_{\text{CoP}}|^{A} = |{\bar y(t + t_{A})}_{\text{CoP}}-{y(t)}_{\text{CoP}}| & \text{for ML.}
    \end{cases}
    \label{eq:anticip_error}
\end{equation}
Considering the reference CoP $(\bar x, \bar y)$ at a subsequent instant $t+t_{A}$, an anticipatory guidance can be provided. When this \textit{anticipatory error} overcome the selected threshold $th_{\text{CoP}}$, the ErgoTac-Belt is triggered and the vibrotactile stimuli is provided in the corresponding direction. 

\begin{algorithm}
\caption{Directional Vibrotactile Feedback}
\label{alg:feedback}
\begin{algorithmic}
\State $th_{\text{CoP}} \leftarrow$ select\_CoP\_threshold;
\State $t_{A} \leftarrow$ select\_anticipatory\_time\_interval ;
\While {true}
    \State $|{\delta x}_{\text{CoP}}|^{A} \leftarrow$ compute\_AP\_error;
    \State $|{\delta y}_{\text{CoP}}|^{A} \leftarrow$ compute\_ML\_error;
    \If{$|{\delta x}_{\text{CoP}}|^{A} > th_{\text{CoP}}$}
            \State active\_ErgoTac-Belt $\rightarrow$ \textit{back};
    \ElsIf{$|{\delta x}_{\text{CoP}}|^{A}  <  - th_{\text{CoP}}$}
            \State active\_ErgoTac-Belt $\rightarrow$ \textit{front};
    \EndIf
    \If{$|{\delta y}_{\text{CoP}}|^{A}  > th_{\text{CoP}}$}
            \State active\_ErgoTac-Belt $\rightarrow$ \textit{left};
    \ElsIf{$|{\delta y}_{\text{CoP}}|^{A}  <  - th_{\text{CoP}}$}
            \State active\_ErgoTac-Belt $\rightarrow$ \textit{right};
    \EndIf
\EndWhile
\end{algorithmic}
\end{algorithm}

\subsection{Performance indicators}

For the sake of validation, also the \textit{real error}, both in the AP and ML axes, was computed as 
\begin{equation}
    \begin{cases}
        |{\delta x}_{\text{CoP}}|^{R} = |{\bar x(t)}_{\text{CoP}}-{x(t)}_{\text{CoP}}| & \text{for AP} \\
        |{\delta y}_{\text{CoP}}|^{R}  = |{\bar y(t)}_{\text{CoP}}-{y(t)}_{\text{CoP}}| & \text{for ML,}
    \end{cases}
    \label{eq:real_error}
\end{equation}
considering the reference CoP $(\bar x, \bar y)$ at the same instant $t$ as the measured CoP $(x,y)$. Based on that, two performance indicators were defined:
\begin{itemize}
    \item \textbf{Root mean square error} $\text{RMSE}(x,y)$ computed as
    \begin{equation}
        \text{RMSE}(x,y) = \sqrt{ \sum_{i=1}^{N} (|{\delta (x,y)}_{\text{CoP}}|^{R})^2/N }
    \label{eq:rmse}
    \end{equation}
    where $N$ is the number of samples of the reference CoP path.
    \item \textbf{Time above threshold} $\text{TAT}(x,y)$ (\%) computed as the percentage of samples for which the condition $|{\delta (x,y)}_{\text{CoP}}|^{R} > th_{\text{CoP}}$ applied, i.e., percentage of time the subjects were far from the reference CoP (the greater TAT, the worse the performance).
\end{itemize}

\section{Experiments}
\label{sec:experiments}

In this section, we present the experimental analysis conducted to assess ErgoTac-Belt capability in guiding the users' CoP along a predefined path. First, the experimental protocol is illustrated. Then, the experimental results are provided. The whole experimental procedure was carried out at Human-Robot Interfaces and Physical Interaction (HRI$^{2}$) Lab, Istituto Italiano di Tecnologia, Genoa, Italy, in accordance with the Declaration of Helsinki, and the protocol was approved by the ethics committee Azienda Sanitaria Locale (ASL) Genovese N.3 (Protocol IIT\_HRII\_ERGOLEAN 156/2020).

\subsection{Experimental protocol}

\begin{figure}[b]
    \centering
    \includegraphics[width = 0.83\columnwidth]{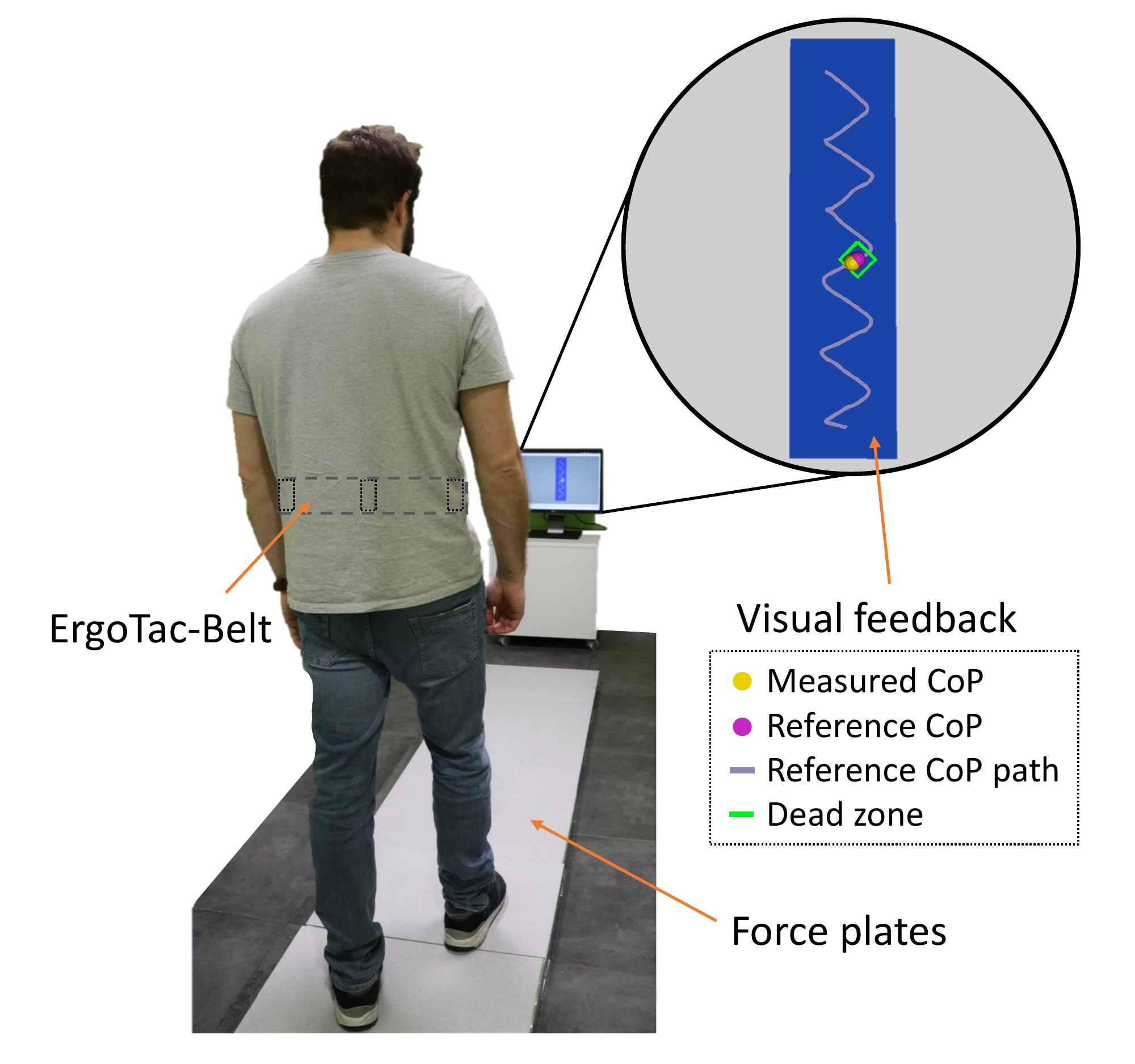}
    \caption{\small Overview of the experimental setup. The visual interface employed for the learning session is highlighted.}
    \label{fig:expsetup}
\end{figure}
Ten healthy human adults ($27.9 \pm 2.1$ years) were recruited in the experiments. Written informed consent was obtained after explaining the experimental procedure. The participants were asked to perform a simple task, i.e., to walk following a reference CoP path, in different experimental conditions. As illustrated in Fig.\ref{fig:expsetup}, six BTS P-6000 force plates were placed in a row to measure the subjects' actual CoP during the task and configured to work as a unique sensor. The reference CoP path was obtained by recording with the force plates the CoP of one of the authors during a slow walk, executed with small steps.
The experiments included two sessions, which entail both a training phase and actual tests, whose conditions and corresponding labels are summarized in Table~\ref{tab:exptrials}. The first session was a learning stage, during which the subjects could learn how to follow a predefined CoP path by reacting to feedback signals. Visual feedback, which is expected to be intuitive and easy to use, was employed for this purpose. The visual interface is depicted in Fig. \ref{fig:expsetup}.
The subjects could see their current (measured) CoP, the reference CoP path outlined on virtual force plates, the reference CoP moving along the path, and a dead zone around the reference CoP that was green or red, when the \textit{real error} $|{\delta (x,y)}_{\text{CoP}}|^{R}$ (eq.\ref{eq:real_error}) was behind or above the threshold $th_{\text{CoP}}$, respectively. First, a training phase was executed, asking the subjects to perform the task using visual feedback. Next, three tests were performed: without feedback and open eyes (NF/O), without feedback and closed eyes, i.e., blind (NF/B), and again with visual feedback (VF/O). The second session was then to test ErgoTac-Belt capability to provide the same guidance as the one offered by visual feedback. In this session, the subjects were asked to wear the ErgoTac-Belt on the torso, directly over the skin, for better perception. First, a training phase was executed. The subjects could do the task with both the visual and ErgoTac-Belt feedback, and then with the ErgoTac-Belt feedback only, both open and closed eyes. Next, two tests were performed: with ErgoTac-Belt feedback and open eyes (EF/O) and with ErgoTac-Belt feedback and closed eyes (EF/B). For the training phases, the subjects chose the number of trials (up to a maximum value) depending on their perceived level of confidence with the task, as not all the subjects have the same sensorimotor skills. Instead, for each test, three trials were performed. For all the subjects, the threshold $th_{\text{CoP}}$ and the anticipatory time interval $t_{A}$ were set to $0.1$\si{\metre} and $0.5$\si{\second}, respectively.
\begin{table}[b]
\caption{Overview of the experimental analysis.}
\begin{tabular}{cccccc} \toprule 
\textbf{Session}            & \textbf{Phase} & \textbf{Feedback} & \textbf{Eyes} & \textbf{N° of trials} & \textbf{Label} \\ \midrule
\multirow{4}{*}{\textbf{1}} & training       & V            & Open          & up to 10              &                \\
                            & test           & No                & Open          & 3                     & NF/O           \\
                            & test           & No                & Closed        & 3                     & NF/B           \\
                            & test           & V            & Open          & 3                     & VF/O           \\ \midrule
\multirow{4}{*}{\textbf{2}} & training       & V + E  & Open          & up to 5               &                \\
                            & training       & E           & Open          & up to 10              &                \\
                            & test           & E           & Open          & 3                     & EF/O           \\
                            & test           & E           & Closed        & 3                     & EF/B  \\ \bottomrule        
\end{tabular}\\ \\
V: Visual; E: ErgoTac-Belt;
\label{tab:exptrials}
\end{table}
The order of the sessions/phases/tests was the same for all the subjects. Since healthy subjects rely on visual feedback during walking, it is needed for learning the task path. Getting familiar with the task directly using the vibrotactile feedback would be extremely hard, if not impossible, even for simple paths (e.g., straight line).

The hypothesis behind this experimental protocol is the following. When the subjects could count on the sensory systems (visual, vestibular, somatosensory) plus the previous knowledge of the environment and task (from learning) and also an additional aid, i.e., the visual feedback (condition VF/O), they likely achieved the best results. 
When removing the additional aid, having the sensory systems and previous knowledge (condition NF/O), they could accomplish the task, but the performance inevitably worsened, even more when removing one of the sensory systems, i.e., the sight (condition NF/B). Our objective here is to investigate whether the vibrotactile feedback, as an additional aid, can bring the performance back/closer to the best one, with or without the sight (condition EF/O and EF/B, respectively).

\subsection{Experimental Results} 
\label{sec:results}

\begin{figure}[bt]
    \centering
    \includegraphics[trim= 10 0 10 15 mm,clip,width=\linewidth]{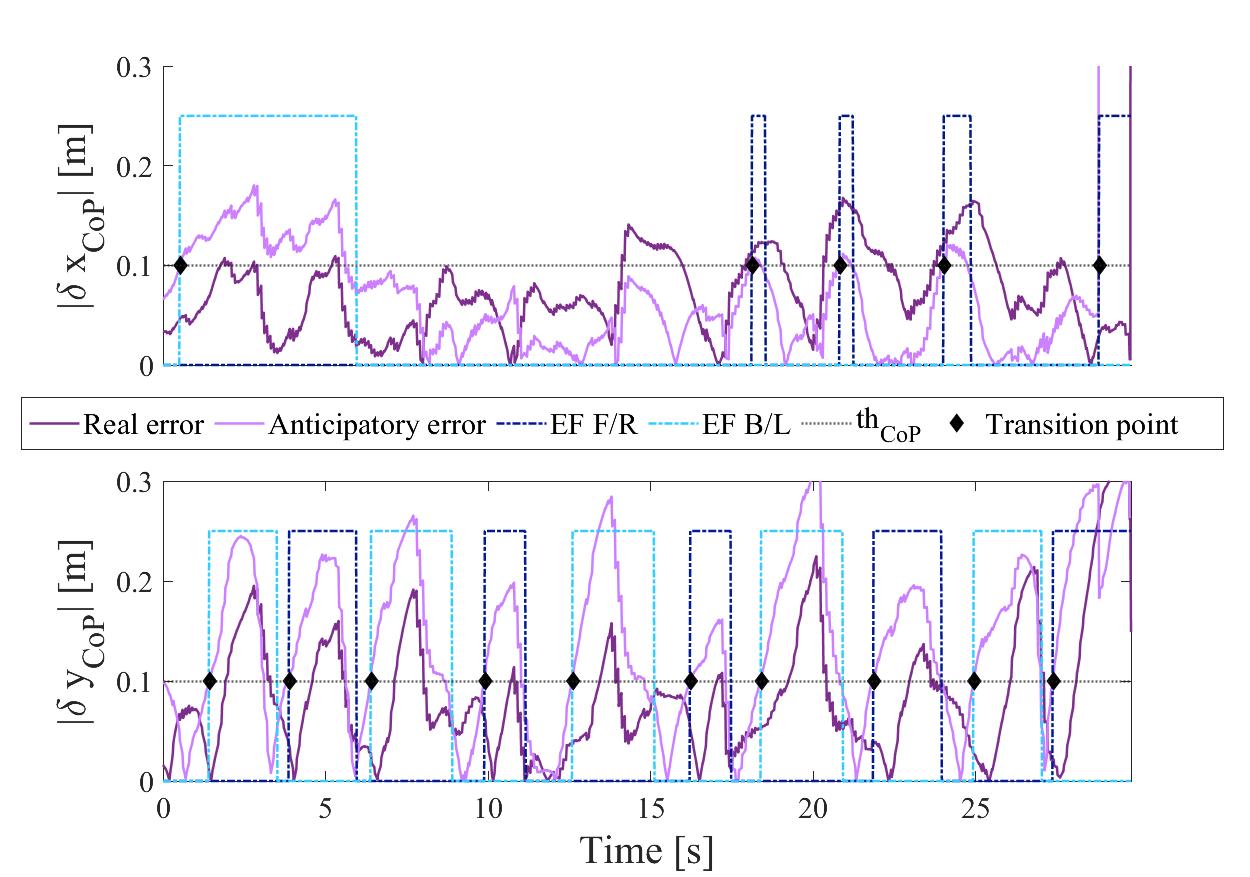}
    \caption{\small Effect of the ErgoTac-Belt feedback (EF) on the $x$ (upper graph) in front (F) and back (B) directions and on the $y$ (lower graph) in right (F) and left (B) directions of the measured CoP error thanks to the anticipatory strategy, for one subject. The transition points where the \textit{anticipatory error} goes above the CoP threshold ($th_{\text{CoP}}$) are also highlighted. 
    }
    \label{fig:ergotac}
    \vspace{-5 mm}
\end{figure}

\begin{figure}[bt]
    \centering
    \includegraphics[trim= 10 0 10 5 mm,clip,width=\linewidth]{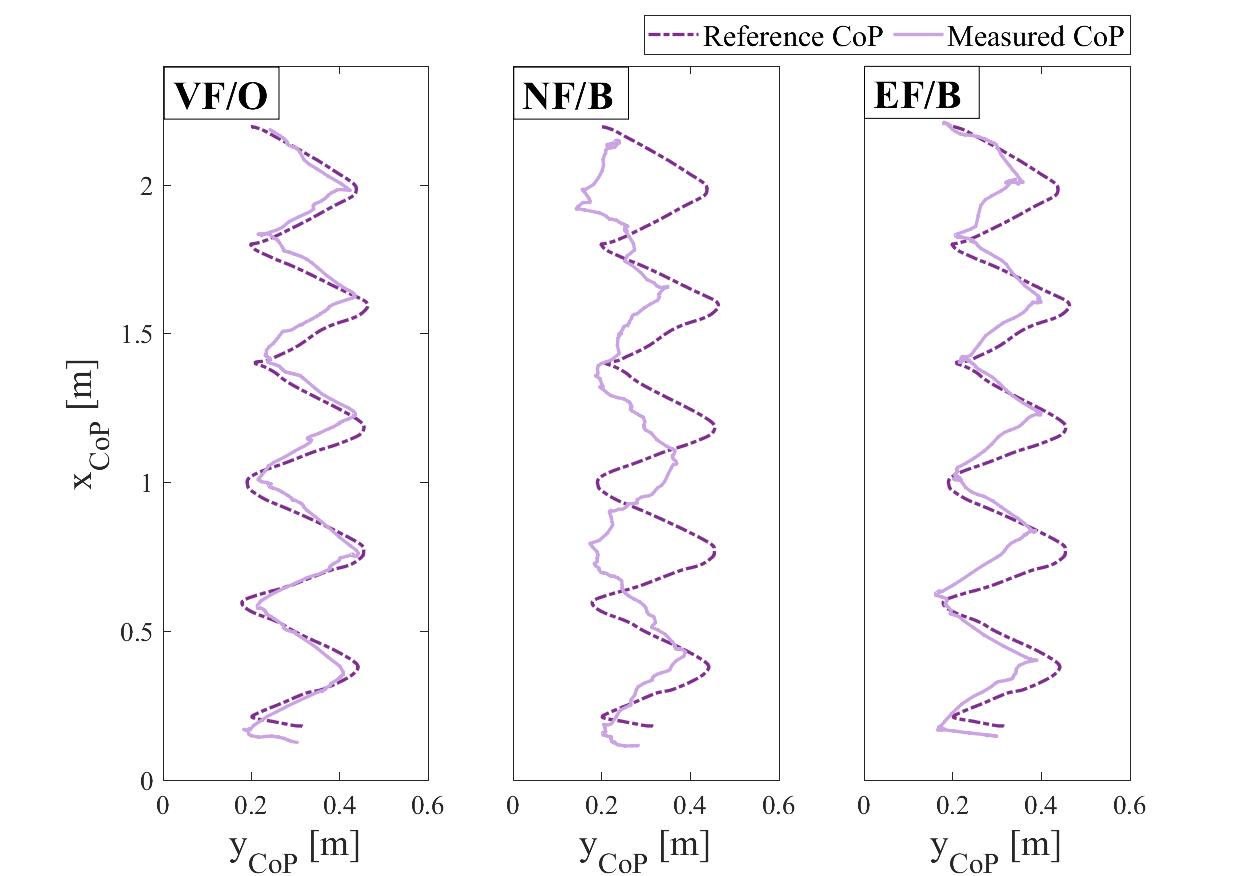}
    \caption{\small CoP reference and measured path in three experimental conditions (see Table), for one subject.}
    \label{fig:compare}
\end{figure}

In Fig.~\ref{fig:ergotac}, the ErgoTac-Belt functioning is illustrated. When the \textit{anticipatory error} $|{\delta (x,y)}_{\text{CoP}}|^{A}$ went above the threshold $th_{\text{CoP}}$ (transition point), in the AP (upper graph) or ML (lower graph) axis, the corresponding vibrotactile unit (F or B, L or R) was activated and the subject reacted to the stimulus, moving in the suggested direction. By using the \textit{anticipatory error} to trigger ErgoTac-Belt, it was possible to keep the \textit{real error} much lower, even below $th_{\text{CoP}}$. The only exception was in the anterior direction. Since the feedback was anticipatory, the stimulus in the F unit (when the subject was slight in advance) was delayed about $t_{A}$. However, he/she could move immediately after to compensate for the \textit{anticipatory error} and quickly decrease the \textit{actual} one.  
Fig.~\ref{fig:compare} depicts the reference CoP path and the measured CoP in the $x,y$ plane during three of the experimental tests, i.e., VF/O, NF/B, and EF/B, respectively, for one subject. It is evident here that the best performance was achieved with the visual feedback (VF/O) and significantly decreased without feedback and closed eyes (NF/B). On the other hand, ErgoTac-Belt feedback, even with closed eyes (EF/B), allowed the measured CoP to come back on track.

\begin{figure}[bt]
    \centering
    \includegraphics[trim= 0 0 0 0 mm,clip,width=\columnwidth]{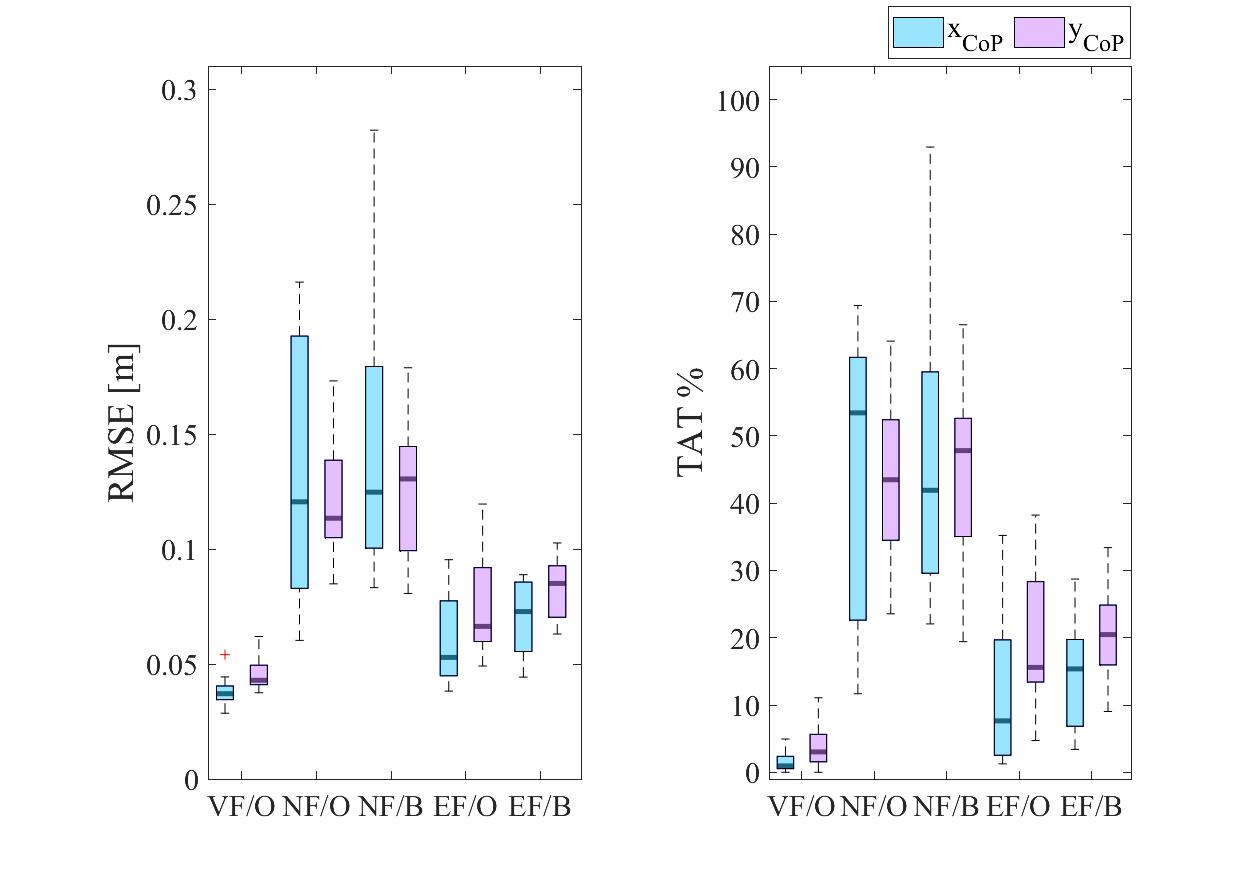}
    \caption{\small Box plots of the RMSE and TAT among all subjects for $x_{\text{CoP}}$ and $y_{\text{CoP}}$ coordinates in the five experimental conditions (Table~\ref{tab:exptrials}).}
    \label{fig:boxplot}
    \vspace{-5 mm}
\end{figure}

Fig.~\ref{fig:boxplot} represents the boxplots of the RMSE$(x,y)$ and TAT$(x,y)$, averaged between the three trials performed, for the five experimental tests, considering all the subjects. As in Fig.~\ref{fig:compare}, it is clear that the best results in accomplishing the task were achieved with the visual feedback (VF/O), where both the RMSE and TAT are extremely low. When removing the feedback, the indicators considerably increase with open eyes (NF/O) and even more with closed eyes (NF/B). 
With the ErgoTac-Belt feedback, conversely, both the RMSE$(x,y)$ and TAT$(x,y)$ were markedly reduced. 
In Table~\ref{tab:rmse} and~\ref{tab:tat}, the RMSE$(x,y)$ and TAT$(x,y)$ for all the subjects in all the tests, averaged among the three trials, are presented. Interestingly, with closed eyes (EF/B) the performance for some subjects (highlighted in grey) was better than with open eyes (EF/O).

\begin{table}[h]
\caption{RMSE$(x,y)$ [cm] for each subject in all the tests.}
\label{tab:rmse}
\begin{center}
\begin{adjustbox}{max width=0.49\textwidth}
\begin{tabular}{|c|c|cc|cc|cc|cc|cc|}
\hline
\multicolumn{2}{|c|}{\multirow{2}{*}{}}             & \multicolumn{2}{c|}{\textbf{VF/O}}            & \multicolumn{2}{c|}{\textbf{NF/O}}              & \multicolumn{2}{c|}{\textbf{NF/B}}              & \multicolumn{2}{c|}{\textbf{EF/O}}             & \multicolumn{2}{c|}{\textbf{EF/B}}             \\ \cline{3-12} 
\multicolumn{2}{|c|}{}                              & \multicolumn{1}{c|}{$x$}      & $y$      & \multicolumn{1}{c|}{$x$}       & $y$       & \multicolumn{1}{c|}{$x$}       & $y$       & \multicolumn{1}{c|}{$x$}      & $y$       & \multicolumn{1}{c|}{$x$}      & $y$       \\ \hline
\multirow{10}{*}{\rotatebox[origin=c]{90}{\textbf{Subjects}}}  & 1 & 2.9 & 4.7 & 19.3 & 14.0 & 17.9 & 11.6 & 3.8 & 6.3 & 5.0 & 7.0 \\
                          & 2 & 3.4 & 6.2 & 12.5 & 11.0 & 10.8 & 14.5 & 7.0 & \cellcolor[HTML]{EFEFEF}10.4 & 7.3 & \cellcolor[HTML]{EFEFEF}10.3 \\
                          & 3 & 4.0 & 4.1 & 21.6 & 8.5 & 10.0 & 15.1 & \cellcolor[HTML]{EFEFEF}9.5 & \cellcolor[HTML]{EFEFEF}12.0 & \cellcolor[HTML]{EFEFEF}8.7 & \cellcolor[HTML]{EFEFEF}8.4 \\
                          & 4 & 3.4 & 3.9 & 8.3 & 11.2 & 14.1 & 13.5 & \cellcolor[HTML]{EFEFEF}7.7 & 5.9 & \cellcolor[HTML]{EFEFEF}7.3 & 6.3 \\
                          & 5 & 3.7 & 4.9 & 15.6 & 13.9 & 24.9 & 12.7 & \cellcolor[HTML]{EFEFEF}5.8 & 6.0 & \cellcolor[HTML]{EFEFEF}5.5 & 6.9  \\
                          & 6 & 4.4 & 5.9 & 19.7 & 17.3 & 28.2 & 17.9 & \cellcolor[HTML]{EFEFEF}8.7 & 9.2 & \cellcolor[HTML]{EFEFEF}4.4 & 9.6 \\
                          & 7 & 3.7 & 4.2 & 11.6 & 11.5 & 10.6 & 9.9 & 4.4 & 4.9 & 6.1 & 9.3 \\
                          & 8 &3.8 & 3.8 & 6.0 & 8.9 & 9.0 & 13.4 & 4.5 & 6.1 & 8.6 & 7.3 \\
                          & 9 & 3.5 & 4.4 & 7.8 & 12.9 & 8.3 & 8.1 & 4.5 & 7.7 & 8.9 & 8.8 \\
                          & 10 & 5.4 & 4.2 & 9.4 & 10.5 & 14.4 & 8.7 & 4.8 & 7.0 & 7.6 & 8.6 \\\hline
\end{tabular}
\end{adjustbox}
\end{center}
\end{table}

\begin{table}[h]
\caption{TAT$(x,y)$ (\%) for each subject in all the tests.}
\label{tab:tat}
\begin{center}
\begin{adjustbox}{max width=0.49\textwidth}
\begin{tabular}{|c|c|cc|cc|cc|cc|cc|}
\hline
\multicolumn{2}{|c|}{\multirow{2}{*}{}}             & \multicolumn{2}{c|}{\textbf{VF/O}}            & \multicolumn{2}{c|}{\textbf{NF/O}}              & \multicolumn{2}{c|}{\textbf{NF/B}}              & \multicolumn{2}{c|}{\textbf{EF/O}}             & \multicolumn{2}{c|}{\textbf{EF/B}}             \\ \cline{3-12} 
\multicolumn{2}{|c|}{}                              & \multicolumn{1}{c|}{$x$}      & $y$      & \multicolumn{1}{c|}{$x$}       & $y$       & \multicolumn{1}{c|}{$x$}       & $y$       & \multicolumn{1}{c|}{$x$}      & $y$       & \multicolumn{1}{c|}{$x$}      & $y$       \\ \hline
\multirow{10}{*}{\rotatebox[origin=c]{90}{\textbf{Subjects}}}  & 1 & 0.0 & 0.0 & 61.7 & 46.1 & 59.5 & 39.1 & 2.5 & 14.2 & 6.2 & 16.0 \\
                          & 2 & 0.7 & 6.6 & 54.7 & 34.5 & 41.8 & 51.1 & 14.8 & 32.2 & 19.0 & 33.4 \\
                          & 3 & 3.6 & 3.1 & 69.4 & 23.6 & 29.6 & 52.6 & \cellcolor[HTML]{EFEFEF}35.2 & \cellcolor[HTML]{EFEFEF}38.2 & \cellcolor[HTML]{EFEFEF}28.7 & \cellcolor[HTML]{EFEFEF}19.6 \\
                          & 4 & 0.5 & 0.3 & 22.6 & 40.9 & 49.7 & 54.6 & \cellcolor[HTML]{EFEFEF}19.7 & \cellcolor[HTML]{EFEFEF}9.1 & \cellcolor[HTML]{EFEFEF}15.7 & \cellcolor[HTML]{EFEFEF}9.0 \\
                          & 5 & 0.5 & 5.6 & 67.6 & 52.4 & 93.0 & 47.4 & \cellcolor[HTML]{EFEFEF}9.6 &\cellcolor[HTML]{EFEFEF} 13.7 & \cellcolor[HTML]{EFEFEF}6.8 & \cellcolor[HTML]{EFEFEF}12.8 \\ 
                          & 6 & 2.4 & 11.1 & 55.2 & 64.1 & 72.2 & 66.5 & \cellcolor[HTML]{EFEFEF}19.7 & 28.3 & \cellcolor[HTML]{EFEFEF}3.4 & 31.3 \\
                          & 7 & 1.2 & 3.0 & 52.2 & 50.0 & 36.9 & 35.1 & 1.3 & 4.7 & 11.7 & 24.9 \\
                          & 8 & 1.7 & 1.6 & 11.7 & 27.4 & 27.1 & 48.3 & 1.3 & 13.4 & 26.2 & 19.4\\
                          & 9 & 0.2 & 3.2 & 21.0 & 52.4 & 22.1 & 19.4 & 4.5 & 20.8 & 18.9 & 23.5 \\
                          & 10 & 4.9 & 2.0 & 26.5 & 36.0 & 42.0 & 25.4 & 5.7 & 17.0 & 15.1 & 21.3 \\ \hline
\end{tabular}
\end{adjustbox}
\end{center}
\vspace{-5 mm}
\end{table}

\section{Discussion}

As demonstrated by the experimental results, when using the ErgoTac-Belt, the subjects were generally able to follow the reference CoP path with good accuracy. The RMSE and TAT, in conditions EF/O and EF/B, were not as low as with visual feedback (VF/O) but were significantly reduced wrt the conditions without any feedback (NF/O and NF/B). The visual feedback, as expected, was more intuitive and easy to use and required a lower reaction time. In fact, the color of the dead zone, depicted in the visual interface (see Fig. \ref{fig:expsetup}), changed the color based on the \textit{real error}, and the subjects were able to promptly respond and maintain the measured CoP close to the reference one. Nevertheless, visual feedback interfaces present several drawbacks: they engage one of the human sensory systems, require a screen or wearable devices which may be demanding in the long run (e.g., augmented reality headsets), and are not suitable for people with diminished or even absent sight. The ErgoTac-Belt, on the other hand, can be worn for a prolonged time without hindering or annoying the users, and its capability in leading the human CoP has proven to be comparable to the one of the visual interface. 
It should be recalled that for ErgoTac-Belt we implemented anticipatory feedback and the same anticipatory time interval $t_A$ was selected for all the subjects. However, as reported by the subjects themselves, a subject-specific threshold may be beneficial to enhance the usability of the vibrotactile device. Given the promising capabilities of ErgoTac-Belt with healthy subjects, the proposed device may be used to guide the CoP of elderly or individuals with vestibular impairments, who may not be aware of or, able to figure out, a safe and ergonomic CoP path during walking. In this paper, the human CoP was measured with a set of force plates. However, by combining motion-capture systems (wearable or camera-based, depending on the application) with online estimation algorithms, it is possible to obtain even a subject-specific model of the whole-body CoP \cite{gonzalez2021extended}. In this way, the ErgoTac-Belt could be used freely in the available space, enabling its applicability in real-life scenarios.   

\section{Conclusions}
In this paper, we presented an anticipatory vibrotactile feedback strategy to lead the human CoP during walking. The ErgoTac-Belt, as an extension of a previous prototype~\cite{kim2018ergotac,kim2021directional}, was proposed to instruct the users about the direction to take via vibrotactile stimuli, both in the AP and ML axes. Experimental results demonstrated its promising capability in guiding healthy subjects along a reference CoP path, with a performance that is comparable to the one achieved by using visual feedback. Future studies will focus on testing the proposed method and device with individuals having vestibular impairments or reduced balancing capability. The subject-specific tuning of the anticipatory strategy, the CoP dead zone shape, and vibrotactile stimuli will be also investigated.

\bibliographystyle{ieeetr}
\bibliography{biblio}

\begin{thebibliography}{10}

\bibitem{horak1996postural}
F.~B. Horak, ``Postural orientation and equilibrium,'' {\em Handbook of
  Physiology, Section 12: Exercise}, pp.~255--292, 1996.

\bibitem{melzer2004postural}
I.~Melzer, N.~Benjuya, and J.~Kaplanski, ``Postural stability in the elderly: a
  comparison between fallers and non-fallers,'' {\em Age and ageing}, vol.~33,
  no.~6, pp.~602--607, 2004.

\bibitem{rubenstein2002epidemiology}
L.~Z. Rubenstein and K.~R. Josephson, ``The epidemiology of falls and
  syncope,'' {\em Clinics in geriatric medicine}, vol.~18, no.~2, pp.~141--158,
  2002.

\bibitem{tinetti2003preventing}
M.~E. Tinetti, ``Preventing falls in elderly persons,'' {\em New England
  journal of medicine}, vol.~348, no.~1, pp.~42--49, 2003.

\bibitem{van2019robotic}
P.~Van~Lam and Y.~Fujimoto, ``A robotic cane for balance maintenance
  assistance,'' {\em IEEE Transactions on Industrial Informatics}, vol.~15,
  no.~7, pp.~3998--4009, 2019.

\bibitem{di2015fall}
P.~Di, Y.~Hasegawa, S.~Nakagawa, K.~Sekiyama, T.~Fukuda, J.~Huang, and
  Q.~Huang, ``Fall detection and prevention control using walking-aid cane
  robot,'' {\em IEEE/ASME Transactions on Mechatronics}, vol.~21, no.~2,
  pp.~625--637, 2015.

\bibitem{ruiz2021improving}
F.~J. Ruiz-Ruiz, A.~Giammarino, M.~Lorenzini, J.~M. Gandarias, J.~M. Gomez-de
  Gabriel, and A.~Ajoudani, ``Improving standing balance performance through
  the assistance of a mobile collaborative robot,'' {\em arXiv preprint
  arXiv:2109.12038}, 2021.

\bibitem{sienko2017role}
K.~Sienko, S.~Whitney, W.~Carender, and C.~Wall~III, ``The role of sensory
  augmentation for people with vestibular deficits: real-time balance aid
  and/or rehabilitation device?,'' {\em Journal of Vestibular Research},
  vol.~27, no.~1, pp.~63--76, 2017.

\bibitem{wall2001balance}
C.~Wall, M.~S. Weinberg, P.~B. Schmidt, and D.~E. Krebs, ``Balance prosthesis
  based on micromechanical sensors using vibrotactile feedback of tilt,'' {\em
  IEEE Transactions on Biomedical Engineering}, vol.~48, no.~10,
  pp.~1153--1161, 2001.

\bibitem{ma2016balance}
C.~Z.-H. Ma, D.~W.-C. Wong, W.~K. Lam, A.~H.-P. Wan, and W.~C.-C. Lee,
  ``Balance improvement effects of biofeedback systems with state-of-the-art
  wearable sensors: A systematic review,'' {\em Sensors}, vol.~16, no.~4,
  p.~434, 2016.

\bibitem{tannert2021immediate}
I.~Tannert, K.~H. Schulleri, Y.~Michel, S.~Villa, L.~Johannsen,
  J.~Hermsd{\"o}rfer, and D.~Lee, ``Immediate effects of vibrotactile
  biofeedback instructions on human postural control,'' in {\em 2021 43rd
  Annual International Conference of the IEEE Engineering in Medicine \&
  Biology Society (EMBC)}, pp.~7426--7432, IEEE, 2021.

\bibitem{ballardini2020vibrotactile}
G.~Ballardini, V.~Florio, A.~Canessa, G.~Carlini, P.~Morasso, and M.~Casadio,
  ``Vibrotactile feedback for improving standing balance,'' {\em Frontiers in
  bioengineering and biotechnology}, vol.~8, p.~94, 2020.

\bibitem{xu2017configurable}
J.~Xu, T.~Bao, U.~H. Lee, C.~Kinnaird, W.~Carender, Y.~Huang, K.~H. Sienko, and
  P.~B. Shull, ``Configurable, wearable sensing and vibrotactile feedback
  system for real-time postural balance and gait training: proof-of-concept,''
  {\em Journal of neuroengineering and rehabilitation}, vol.~14, no.~1,
  pp.~1--10, 2017.

\bibitem{nanhoe2012effects}
W.~Nanhoe-Mahabier, J.~Allum, E.~Pasman, S.~Overeem, and B.~Bloem, ``The
  effects of vibrotactile biofeedback training on trunk sway in parkinson's
  disease patients,'' {\em Parkinsonism \& related disorders}, vol.~18, no.~9,
  pp.~1017--1021, 2012.

\bibitem{sienko2012biofeedback}
K.~H. Sienko, M.~D. Balkwill, and C.~Wall, ``Biofeedback improves postural
  control recovery from multi-axis discrete perturbations,'' {\em Journal of
  neuroengineering and rehabilitation}, vol.~9, no.~1, pp.~1--11, 2012.

\bibitem{haggerty2012effects}
S.~Haggerty, L.-T. Jiang, A.~Galecki, and K.~H. Sienko, ``Effects of
  biofeedback on secondary-task response time and postural stability in older
  adults,'' {\em Gait \& posture}, vol.~35, no.~4, pp.~523--528, 2012.

\bibitem{alahakone2010real}
A.~U. Alahakone and S.~A. Senanayake, ``A real-time system with assistive
  feedback for postural control in rehabilitation,'' {\em IEEE/ASME
  Transactions on Mechatronics}, vol.~15, no.~2, pp.~226--233, 2010.

\bibitem{kentala2003reduction}
E.~Kentala, J.~Vivas, and C.~Wall~III, ``Reduction of postural sway by use of a
  vibrotactile balance prosthesis prototype in subjects with vestibular
  deficits,'' {\em Annals of Otology, Rhinology \& Laryngology}, vol.~112,
  no.~5, pp.~404--409, 2003.

\bibitem{verhoeff2009effects}
L.~L. Verhoeff, C.~G. Horlings, L.~J. Janssen, S.~A. Bridenbaugh, and J.~H.
  Allum, ``Effects of biofeedback on trunk sway during dual tasking in the
  healthy young and elderly,'' {\em Gait \& posture}, vol.~30, no.~1,
  pp.~76--81, 2009.

\bibitem{janssen2010salient}
M.~Janssen, R.~Stokroos, J.~Aarts, R.~van Lummel, and H.~Kingma, ``Salient and
  placebo vibrotactile feedback are equally effective in reducing sway in
  bilateral vestibular loss patients,'' {\em Gait \& posture}, vol.~31, no.~2,
  pp.~213--217, 2010.

\bibitem{ma2015vibrotactile}
C.~Z.-H. Ma, A.~H.-P. Wan, D.~W.-C. Wong, Y.-P. Zheng, and W.~C.-C. Lee, ``A
  vibrotactile and plantar force measurement-based biofeedback system: Paving
  the way towards wearable balance-improving devices,'' {\em Sensors}, vol.~15,
  no.~12, pp.~31709--31722, 2015.

\bibitem{ma2017wearable}
C.~Z.-H. Ma and W.~C.-C. Lee, ``A wearable vibrotactile biofeedback system
  improves balance control of healthy young adults following perturbations from
  quiet stance,'' {\em Human movement science}, vol.~55, pp.~54--60, 2017.

\bibitem{afzal2015portable}
M.~R. Afzal, M.-K. Oh, C.-H. Lee, Y.~S. Park, and J.~Yoon, ``A portable gait
  asymmetry rehabilitation system for individuals with stroke using a
  vibrotactile feedback,'' {\em BioMed research international}, vol.~2015,
  2015.

\bibitem{crea2014providing}
S.~Crea, C.~Cipriani, M.~Donati, M.~C. Carrozza, and N.~Vitiello, ``Providing
  time-discrete gait information by wearable feedback apparatus for lower-limb
  amputees: usability and functional validation,'' {\em IEEE Transactions on
  Neural Systems and Rehabilitation Engineering}, vol.~23, no.~2, pp.~250--257,
  2014.

\bibitem{winter1995abc}
D.~A. Winter, {\em ABC (anatomy, biomechanics and control) of balance during
  standing and walking}.
\newblock Waterloo Biomechanics, 1995.

\bibitem{lee2012directional}
B.-C. Lee, B.~J. Martin, and K.~H. Sienko, ``Directional postural responses
  induced by vibrotactile stimulations applied to the torso,'' {\em
  Experimental brain research}, vol.~222, no.~4, pp.~471--482, 2012.

\bibitem{vuillerme2007controlling}
N.~Vuillerme, O.~Chenu, J.~Demongeot, and Y.~Payan, ``Controlling posture using
  a plantar pressure-based, tongue-placed tactile biofeedback system,'' {\em
  Experimental brain research}, vol.~179, no.~3, pp.~409--414, 2007.

\bibitem{fani2021multi}
S.~Fani, S.~Ciotti, and M.~Bianchi, ``Multi-cue haptic guidance through
  wearables for enhancing human ergonomics,'' {\em IEEE Transactions on
  Haptics}, 2021.

\bibitem{kinnaird2016effects}
C.~Kinnaird, J.~Lee, W.~J. Carender, M.~Kabeto, B.~Martin, and K.~H. Sienko,
  ``The effects of attractive vs. repulsive instructional cuing on balance
  performance,'' {\em Journal of neuroengineering and rehabilitation}, vol.~13,
  no.~1, pp.~1--5, 2016.

\bibitem{Shull2015}
P.~B. Shull and D.~Damian, ``Haptic wearables as sensory replacement, sensory
  augmentation and trainer--a review,'' {\em Journal of NeuroEngineering and
  Rehabilitation}, vol.~12, 2015.

\bibitem{tan2019user}
J.~Tan, Y.~Ge, X.~Sun, Y.~Zhang, and Y.~Liu, ``User experience of tactile
  feedback on a smartphone: Effects of vibration intensity, times and
  interval,'' in {\em International Conference on Human-Computer Interaction},
  pp.~397--406, Springer, 2019.

\bibitem{filosa2018new}
M.~Filosa, I.~Cesini, E.~Martini, G.~Spigler, N.~Vitiello, C.~Oddo, and
  S.~Crea, ``A new sensory feedback system for lower-limb amputees: Assessment
  of discrete vibrotactile stimuli perception during walking,'' in {\em
  International Symposium on Wearable Robotics}, pp.~105--109, Springer, 2018.

\bibitem{crea2017time}
S.~Crea, B.~B. Edin, K.~Knaepen, R.~Meeusen, and N.~Vitiello, ``Time-discrete
  vibrotactile feedback contributes to improved gait symmetry in patients with
  lower limb amputations: Case series,'' {\em Physical therapy}, vol.~97,
  no.~2, pp.~198--207, 2017.

\bibitem{Li2015}
Y.~Li, W.~R. Jeon, and C.~S. Nam, ``Navigation by vibration: Effects of
  vibrotactile feedback on a navigation task,'' {\em International Journal of
  Industrial Ergonomics}, vol.~46, pp.~76 -- 84, 2015.

\bibitem{kim2021directional}
W.~Kim, V.~R. Garate, J.~M. Gandarias, M.~Lorenzini, and A.~Ajoudani, ``A
  directional vibrotactile feedback interface for ergonomic postural
  adjustment,'' {\em IEEE Transactions on Haptics}, 2021.

\bibitem{brewster2004tactons}
S.~A. Brewster and L.~M. Brown, ``Tactons: structured tactile messages for
  non-visual information display,'' 2004.

\bibitem{kim2018ergotac}
W.~Kim, M.~Lorenzini, K.~Kap{\i}c{\i}o{\u{g}}lu, and A.~Ajoudani, ``{ErgoTac:
  A} tactile feedback interface for improving human ergonomics in workplaces,''
  {\em IEEE Robotics and Automation Letters}, vol.~3, no.~4, pp.~4179--4186,
  2018.

\bibitem{gonzalez2021extended}
A.~Gonz{\'a}lez, P.~Fraisse, and M.~Hayashibe, ``An extended statically
  equivalent serial chain—identification of whole body center of mass with
  dynamic motion,'' {\em Gait \& Posture}, vol.~84, pp.~45--51, 2021.

\end{thebibliography}

\addtolength{\textheight}{-12cm}   

\end{document}